\documentclass[conference]{IEEEtran}

\IEEEoverridecommandlockouts

\usepackage{cite}
\usepackage[absolute,overlay]{textpos}
\usepackage{fancyhdr}
\usepackage{amsmath,amssymb,amsfonts}
\usepackage{algorithmic}
\usepackage{graphicx}
\usepackage{textcomp}
\usepackage{xcolor}
\usepackage{algorithm}
\def\BibTeX{{\rm B\kern-.05em{\sc i\kern-.025em b}\kern-.08em
    T\kern-.1667em\lower.7ex\hbox{E}\kern-.125emX}}

\begin{document}
\newif\ifanonymous
\anonymousfalse 

\title{Where to Explore: A Reach and Cost-Aware Approach for Unbiased Data Collection in Recommender Systems}
\ifanonymous

\author{
  \IEEEauthorblockN{Anonymous Author(s)}
  \IEEEauthorblockA{
    Affiliation\\
    Email address
  }
}

\else
\author{\IEEEauthorblockN{1\textsuperscript{st} Qiang Chen}
\IEEEauthorblockA{\textit{Tubi}\\
San Francisco, United States \\
qiang@tubi.tv}
\and
\IEEEauthorblockN{2\textsuperscript{nd} Venkatesh Ganapati Hegde}
\IEEEauthorblockA{\textit{Tubi}\\
San Francisco, United States \\
vhegde@tubi.tv}
}
\fi

\maketitle

\begin{textblock*}{\textwidth}(0cm,26cm) 
  \centering
  \footnotesize
  \copyright~2025 IEEE. Personal use of this material is permitted. Permission from IEEE must be obtained for all other uses, in any current or future media, including reprinting/republishing this material for advertising or promotional purposes, creating new collective works, for resale or redistribution to servers or lists, or reuse of any copyrighted component of this work in other works.
\end{textblock*}

\pagestyle{fancy}
\fancyhf{} 
\fancyfoot[C]{\footnotesize \copyright~2025 IEEE. Personal use of this material is permitted. 
Permission from IEEE must be obtained for all other uses, in any current or future media, 
including reprinting/republishing this material for advertising or promotional purposes, 
creating new collective works, for resale or redistribution to servers or lists, 
or reuse of any copyrighted component of this work in other works.}

\setlength{\footskip}{20pt}  

\begin{abstract}

Exploration is essential to improve long-term recommendation quality, but it often degrades short-term business performance, especially in remote-first TV environments where users engage passively, expect instant relevance, and offer few chances for correction. This paper introduces an approach for delivering content-level exploration safely and efficiently by optimizing its placement based on reach and opportunity cost. 
Deployed on a large-scale streaming platform with over 100 million monthly active users, our approach identifies scroll-depth regions with lower engagement and strategically introduces a dedicated container, the \textit{"Something Completely Different"} row containing randomized content. Rather than enforcing exploration uniformly across the user interface (UI), we condition its appearance on empirically low-cost, high-reach positions to ensure minimal tradeoff against platform-level watch time goals. 
Extensive A/B testing shows that this strategy preserves business metrics while collecting unbiased interaction data. Our method complements existing intra-row diversification and bandit-based exploration techniques by introducing a deployable, behaviorally informed mechanism for surfacing exploratory content at scale. Moreover, we demonstrate that the collected unbiased data, integrated into downstream candidate generation, significantly improves user engagement, validating its value for recommender systems.
\end{abstract}


\section{Introduction}

Recommender systems play a central role in helping users navigate increasingly large and personalized content catalogs. These systems typically learn from user interactions, such as plays, dwell time, and watch duration, to personalize rankings over time \cite{davidson2010youtube, zhang2020recommendation, chen2019top}. However, because this feedback is shaped by what the system has already chosen to present, recommender pipelines are susceptible to feedback loops \cite{yuan2019improving}. One well-documented outcome is presentation bias, in which the visibility of content, rather than its intrinsic relevance, disproportionately drives engagement \cite{chen2023keynote, jiang2019degenerate, ai2018unbiased}.

Exploration is often introduced to reduce this bias by surfacing unfamiliar or underexposed content. However, in practice, exploration is difficult to deploy safely\cite{shaped2023explore,chen2023bias}. In long-form, remote-first streaming environments, users exhibit passive interaction patterns, strong homepage dependence, and limited willingness to navigate away from familiar surfaces. Search usage is low, and users expect relevant content to be surfaced with minimal effort. As a result, introducing randomized content too prominently, or to the wrong audience segment, can negatively impact short-term engagement metrics.

We propose a cost-sensitive, behaviorally informed approach for delivering content-level exploration without degrading core user or business outcomes. Rather than modifying ranking logic or attempting to model user intent directly, our method conditions the \textbf{delivery} of exploratory content on session context, specifically, scroll depth as a proxy for user readiness. When users scroll deeper into the UI, they are more likely to be disengaged from top-ranked results and therefore more receptive to novelty.

Our approach introduces a dedicated, lightly filtered surface for exploration content, triggered only when users reach empirically low-cost, high-reach scroll-depth regions. This allows for randomized data collection at scale without requiring separate navigation flows or changes to interaction paradigms. Extensive A/B testing demonstrates that our method preserves key engagement metrics while capturing high-quality, unbiased user signals. The technique is complementary to intra-row bandit strategies and diversity-aware ranking policies, and offers a safe, modular deployment pathway for platforms seeking to improve long-term recommendation quality without incurring short-term performance risk.

\section{Challenge}

Although exploration is widely acknowledged as essential for long-term recommendation quality, it presents a well-known short-term tradeoff: unfamiliar content often leads to reduced engagement, especially when surfaced too prominently or to the wrong users.

This challenge is particularly acute in remote-first streaming environments. Users typically rely on a small number of visible homepage rows, interact passively via scrolling, and rarely engage with search or secondary navigation. As a result, platforms face extreme pressure to maximize immediate relevance above the fold, leaving little room for error when introducing randomized or exploratory content.

While prior work has explored ranking-based diversification or bandit-based content injection, these approaches often assume uniform deployment across users or surfaces. In reality, the cost of placing exploratory content varies by position, session, and user behavior. Without a mechanism to measure or respect this cost, even well-intentioned exploration can lead to metric regression.

The core challenge, then, is not just what content to explore-but how to deliver exploration in a way that is \textbf{behaviorally aligned}, \textbf{cost-sensitive}, and \textbf{scalable}. Solving this requires identifying regions in the UI where exploration can succeed without degrading core engagement-and doing so in a way that is simple to deploy, monitor, and iterate upon.

\section{Related Work}

Recommender systems rely heavily on user interaction data to learn and refine personalized suggestions. However, this data is inherently biased due to position effects, user behavior patterns, and model-driven exposure feedback loops. To mitigate these limitations, a rich body of work has explored various exploration strategies aimed at collecting more informative, diverse, and representative user feedback. These efforts span model-level algorithms rooted in reinforcement learning and bandit theory, user interface–level interventions that govern how and where novel content is displayed, and behaviorally informed signals used to optimize delivery timing and context. In this section, we survey these complementary threads of research and position our contribution within this broader landscape.

\subsection{Algorithmic Exploration Strategies}

Exploration in recommender systems is often framed through the lens of reinforcement learning, where the system selects one item at a time to maximize long-term reward, and a ranked list is formed by sequentially concatenating these choices. Within this approach, a variety of algorithmic strategies from the reinforcement learning literature have been adapted for use in recommendation contexts.

Epsilon-Greedy: One of the simplest and most widely known approaches is the epsilon-greedy strategy  \cite{sutton2018reinforcement,lattimore2020bandit}. At each decision point, the algorithm chooses a random item with a small probability epsilon (commonly 5\%), and otherwise selects the item with the highest predicted score. This method is easy to implement and integrates seamlessly with modern ranking models, such as deep neural networks. However, since the exploratory choice is uninformed by model uncertainty, it often results in suboptimal recommendations that can negatively affect short-term user experience, because the randomness introduced is not guided by any model uncertainty, leading to suboptimal exploration and unsatisfactory recommendations.

Upper Confidence Bound (UCB): UCB-based strategies \cite{auer2002finite} improve upon epsilon-greedy by incorporating model uncertainty into the exploration decision. These methods assign an exploration bonus to items with higher uncertainty, quantified through confidence intervals around their predicted scores, thereby balancing exploitation (high predicted value) and exploration (low confidence). In the ranking context, this results in items with a combination of high potential and limited historical interaction being promoted. UCB approaches are particularly effective when confidence estimation is reliable, often outperforming epsilon-greedy exploration. However, their effectiveness relies on accurate uncertainty estimation, which constrains the class of models that can be used, limiting compatibility with complex architectures such as gradient-boosted trees or deep learning models without additional confidence estimation mechanisms.

Thompson Sampling: Another principled approach is Thompson Sampling \cite{chapelle2011empirical, thompson1933likelihood}, which models user-item interaction probabilities as posterior distributions. At serving time, scores are sampled from these distributions and items are ranked accordingly. This method naturally balances exploration and exploitation through stochastic sampling. However, integrating Thompson Sampling with modern recommendation architectures, such as gradient-boosted trees or deep learning models, can be challenging. Moreover, its implicit nature makes it difficult to precisely control or quantify the degree of exploration, which complicates its application in production settings with strict business constraints or targeted unbiased data collection goals.

Active Learning\cite{active_learning2015}: Active learning approaches focus on identifying which data points, if labeled or interacted with, would most improve the model. In the context of recommender systems, this typically translates to proactively selecting items to present to users in order to maximize learning efficiency. Rather than relying on random or uncertainty-based sampling, active learning strategies consider the expected model improvement or information gain associated with user feedback on specific items. For instance, models may prioritize presenting items that are most likely to resolve uncertainty in user preference embeddings or disambiguate item clusters. While promising in theory, active learning approaches often assume access to an oracle or labeler and are more commonly studied in small-scale or simulation-based environments. Their integration into large-scale online systems remains limited due to challenges in modeling real-time feedback loops, incorporating business constraints, and ensuring exploration remains behaviorally acceptable to users.

While the above methods offer theoretically grounded mechanisms for exploration, they typically operate at the model level and abstract away from user interface factors, behavioral patterns, and platform-specific business constraints. 

\subsection{UI-Level Interventions for Exploration}

Several large-scale platforms use dedicated UI surfaces to promote discovery outside of a user's typical preferences. These include explore tabs, trending rows, and curated modules that sit alongside core personalized content.

YouTube’s \textit{fixed-position exploration slot}, described in recent work \cite{youtube_fixed_slot}, holds out a single recommendation slot for exploratory content retrieved from a novel source, bypassing the main ranker. While this design enables unbiased interaction measurement, the method for selecting the placement position is not specified, nor is it evaluated in terms of user exposure or engagement tradeoffs.

YouTube also offers a “New to You” tab, which introduces content outside of the user's past viewing history \cite{youtube_newtoyou}. This surface is navigable from the homepage but requires an explicit tab switch, potentially limiting its reach to users who are already in an exploratory mindset.

Netflix maintains a dedicated “Top 10” row showcasing global trending content \cite{netflix_top10}. Although this promotes breadth, it concentrates exposure on popular titles, potentially reinforcing popularity bias and limiting diversity.

These interventions show a growing interest in UI-level exploration, but generally rely on static positioning, heuristic placement, or opt-in user behavior. To our knowledge, no publicly documented system provides a principled approach for deciding where exploration should be delivered within the core recommendation surface

\subsection{Multi-Task Learning and Diversity-Aware Recommendation}

Another relevant line of work focuses on improving recommendation quality through multi-objective modeling and diversity-aware ranking. Multi-task learning approaches jointly predict multiple user intents, such as clicks, purchases, likes, and watch time, and blend these objectives to generate a final ranking score \cite{li2023intent}. These models aim to better capture user preferences across different engagement types and are often used to improve overall ranking robustness.

Related efforts in diversity-aware recommendation seek to expose users to a broader range of content by optimizing for intra-list diversity, genre coverage, or novelty. Intent-aware models and mixture-of-objectives ranking strategies fall into this category, often balancing relevance with exploration-like signals to encourage content discovery \cite{jannach2024survey}.

While these methods can increase exposure to underrepresented items, they are still fundamentally exploitative in nature, as they rely on learned user preferences and deterministic ranking.

\section{Key Contribution}

Our work addresses a critical gap in the design of exploration mechanisms for recommender systems: not just what to explore, but where to surface exploratory content in a way that balances user experience with data collection needs.

While prior systems have introduced explore tabs, fixed recommendation slots, and popular content rows, these interventions are typically static in placement, opt-in by design, or narrowly focused on head content. For instance, YouTube’s recent work on targeted exploration introduces a fixed-position slot to surface novel candidates, but does not describe how that position was selected or evaluated. Other systems, such as a “New to You” tab or global “Top 10” row, promote breadth but rely on user intent or reinforce popularity bias.

Our contribution is a scroll-depth–conditioned, cost-aware approach for delivering exploratory content within the homepage experience. We introduce a principled method for selecting a fixed placement location, one that balances user reach with low opportunity cost,  allowing randomized content to be surfaced without disrupting high-performing regions of the UI.

This approach enables unbiased signal collection at scale, particularly on underexposed content, while preserving short-term engagement. It supports platforms in deploying exploration safely, without relying on personalization, session-based gating, or changes to user behavior.

\section{Proposed Method}

We frame exploration not only as a question of what content to surface, but as a question of how to deliver it in a way that preserves short-term engagement while enabling long-term learning. Our method introduces a cost-aware, behaviorally triggered delivery strategy that supports safe and scalable deployment of randomized content in real-world recommender systems.

\subsection{Cost-Sensitive Delivery Strategy}

\begin{figure}
  \includegraphics[width=\linewidth]{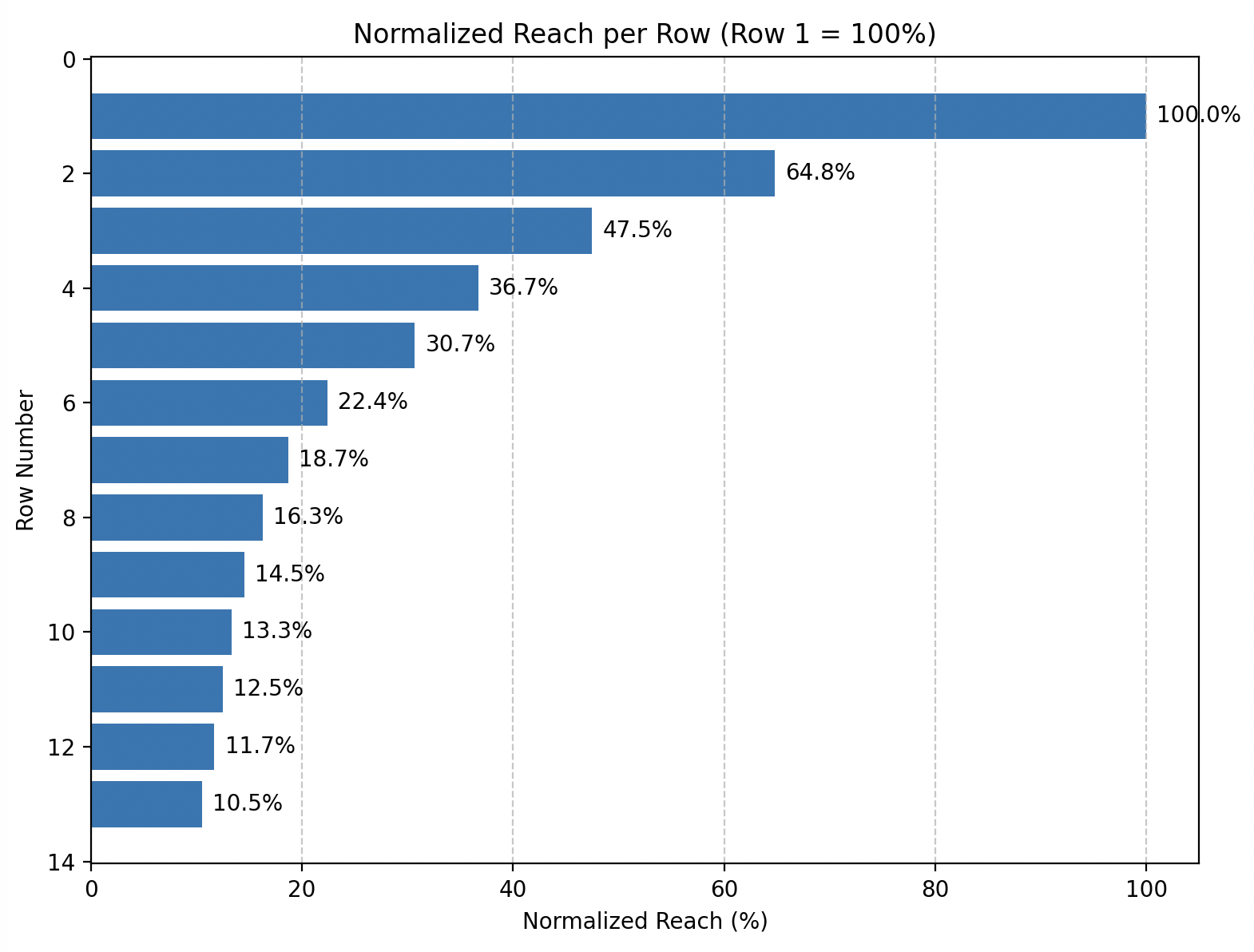}
  \caption{
   Normalized Reach per Row (Row 1 = 100\%), Reach per row, by removing some
   special container e.g. continue watching row. 
  }
  \label{fig:reach}
\end{figure}

In a homepage-style recommender UI, not all row placements carry equal value. Some rows, especially those near the top, contribute significantly to user engagement, while others have broad reach but lower marginal engagement. We define an empirical approach to assess opportunity cost using two factors:
\begin{itemize}
    \item \textbf{Reach}: The proportion of user sessions in which a given row is visible (see Figure~\ref{fig:reach}).
    \item \textbf{User Engagemente Contribution}: The user engagement attributable to that row.
\end{itemize}

In practice, we select a row that meets two criteria: it contributes approximately 1\% of homepage user engagement and is visible in about 10\% of user sessions. This placement strikes a balance between sufficient exposure for data collection and minimal disruption to engagement metrics.

Rather than re-ranking existing containers or dynamically selecting exploration rows per session, we adopt a fixed-placement strategy. A dedicated exploratory row is positioned at a scroll-depth region identified through the cost model as low-risk but high-reach. This approach ensures that exploration is surfaced only when users are behaviorally primed for novelty, enabling the platform to collect unbiased interaction data without prematurely disrupting personalized experiences.

\subsection{Implementation of the \ifanonymous Dedicated Row \else “Something Completely Different” Row \fi}

To implement this delivery strategy, we introduced the \ifanonymous dedicated row\else “Something Completely Different” (SCD) row \fi: a randomized, lightly filtered content container placed near the scroll-depth frontier identified by the cost-reach analysis. It appears only when users reach that region, which we interpret as a behavioral signal of disengagement from top-ranked content.

While it is technically possible to conduct exploration over the entire content catalog, in practice, minimal filtering is applied to ensure user experience remains safe. Specifically, we curate a pool of high-quality, engagement-eligible titles that are suitable for randomized exposure. This lightweight qualification step helps avoid surfacing content with known quality or policy concerns, while still covering a broad and representative portion of the catalog.

 The \ifanonymous dedicated  \else SCD \fi row does not apply re-ranking or personalization logic. Instead, it samples uniformly from the qualified exploration pool. This design allows the system to collect off-policy interaction signals under randomized exposure conditions, which are critical for generating unbiased data.

\subsection{Deployment Guardrails via Controlled Experiments}

Exploration is only surfaced when safe to do so. All placement decisions are validated via A/B tests before rollout. Guardrails ensure:
\begin{itemize}
    \item No significant regression in core metrics (user engagement)
    \item Positive or neutral signal quality in user interactions
\end{itemize}

If these conditions are not met, the exploratory surface is either removed or repositioned. This ensures that the system remains adaptable to context while maintaining engagement.

\ifanonymous
\else
\begin{figure}
  \includegraphics[width=\linewidth]{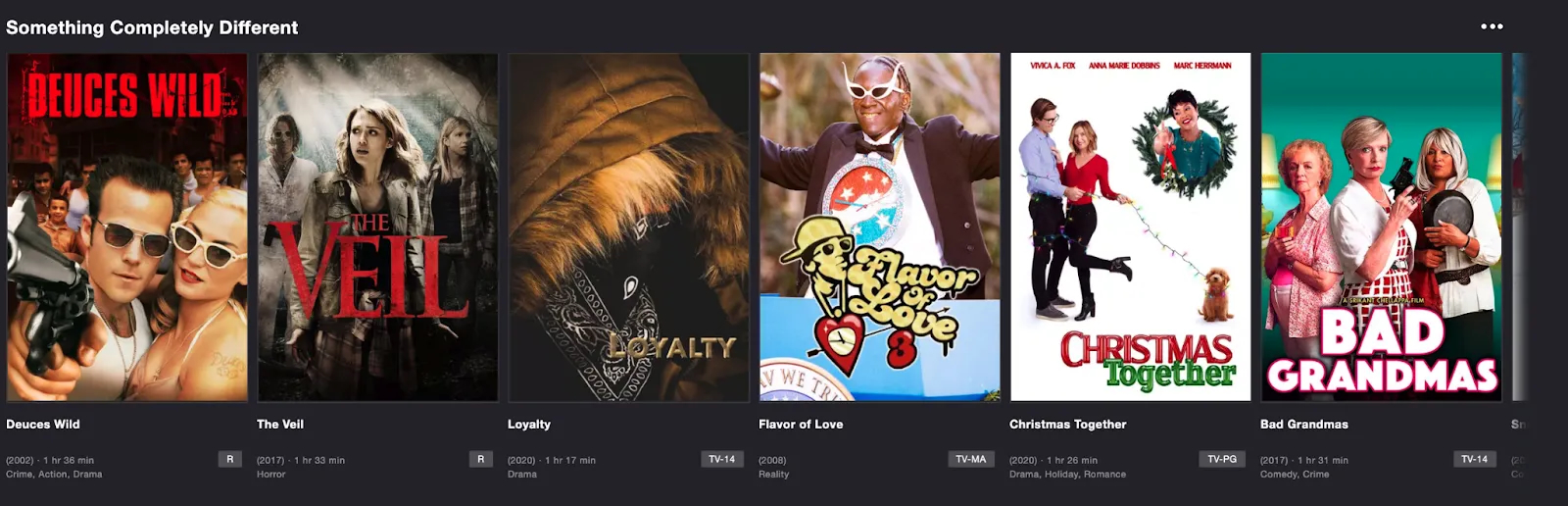}
  \caption{\ifanonymous Dedicated Row \else “Something Completely Different” row, a randomized, lightly filtered content container placed near the scroll-depth frontier identified by the cost-reach analysis. It appears only when users reach that region, which we interpret as a behavioral signal of disengagement from top-ranked content. \fi}
  \label{fig:exploration_row}
\end{figure}
\fi
\section{Experiment Results}

To validate the effectiveness of our proposed delivery strategy, we conducted a comparative experiment between two UI-level placements for exploration content: (1) a \ifanonymous dedicated row\else “Something Completely Different” (SCD) row \fi, and (2) partial insertion of exploratory items into the existing top “Recommended for You” row.

\subsection{Exploration via Insertion into Personalized Row}

In the second strategy, we preserved the personalized content in the “Recommended for You” row while inserting a small number of uniformly selected exploratory titles. The goal was to isolate 1\% of total engagement for exploration, comparable to that of the dedicated row.

The procedure was as follows:

\begin{enumerate}
    \item Identify the “Recommended for You” row for each user session.
    \item Estimate the number of exploratory positions needed to reach 1\% total engagement. For example, if the row contributes 20\% of overall user engagement, then inserting exploration content into 5\% of its positions yields a 1\% overall impact ($5\% \times 20\% = 1\%$).
    \item Select a pool of high-quality, high-diversity titles suitable for exploration.
    \item Randomly sample $N$ titles from this pool, where $N = 5\% \times$ length of the row.
    \item Randomly choose $N$ positions to insert the exploratory titles, increasing the row length accordingly.
\end{enumerate}

\subsection{Results and Discussion}

The experiment was conducted on a large-scale streaming platform with three user groups. The \textbf{Control} group received the standard homepage experience with no exploratory content. In the \textbf{Recommended Row (insertion)} group, a small number of randomly selected titles were inserted into randomly chosen positions within the existing “Recommended” row. In contrast, the \textbf{SCD Row (dedicated)} group was presented with a dedicated “Something Completely Different” row positioned near the bottom of the homepage, designed to surface randomized content as users reached deeper scroll depths. Table~\ref{significance} presents the impact of each treatment on user engagement.

\begin{table}[ht]
\caption{Comparison of Exploration Placements}
\centering
\begin{tabular}{lcc}
\hline
\textbf{Treatment} & \textbf{User Engagement Lift} & \textbf{p-value} \\
\hline
Control (no exploration) & – & – \\
Recommended Row (insertion) & \textminus0.13\% & 0.431 \\
\ifanonymous Dedicated \else SCD \fi Row (dedicated) & +0.28\% & 0.062 \\
\hline
\end{tabular}
\label{significance}
\end{table}

While both treatments exposed users to the same set of randomized titles, their effects on engagement diverged. The dedicated exploratory row produced a modest positive lift of +0.28\%, with a $p$-value of 0.062, just above the conventional threshold for statistical significance. In contrast, the insertion-based strategy within the personalized row yielded a slight negative effect of \textminus0.13\% ($p = 0.431$), suggesting it may have subtly disrupted user expectations.

Although neither result is statistically significant, the observed trend indicates that users may be more receptive to exploratory content when it is surfaced through a dedicated, behaviorally triggered UI component, rather than embedded within core personalized rows. These findings reinforce the importance of thoughtful placement and context in exploration design and suggest that delivery mechanism plays a key role in shaping user response. Further experimentation with larger samples or refined targeting may help validate and build on this trend.

\section{Unbiased data value understanding and utilization}

\subsection{Unbiased data understanding}

\begin{figure}
  \includegraphics[width=\linewidth]{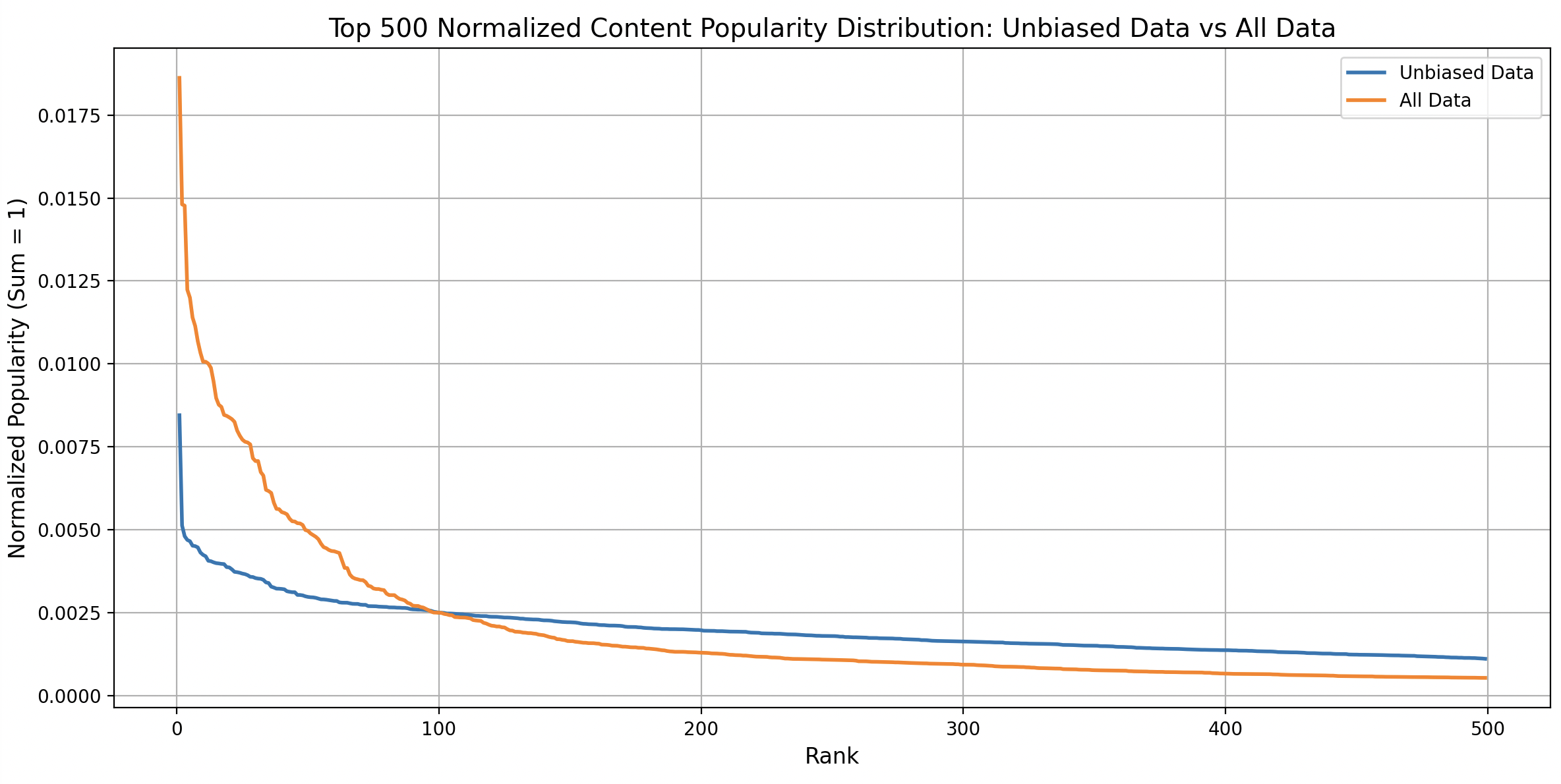}
  \caption{
    Normalized popularity distribution of the top 500 programs: Unbiased vs. Overall Data. 
    Each point on the graph represents a single program. The x-axis denotes the rank (1 = highest), and the y-axis shows the normalized popularity score. 
    Both distributions are normalized to sum to 1, allowing for direct comparison. The flatter curve of the unbiased data reflects a more equitable exposure of content.
  }
  \label{fig:popularity_diff}
\end{figure}

The Figure \ref{fig:popularity_diff} compares the normalized popularity distributions of the top 500 programs from two data sources using one week of data: the unbiased recommendation output and the overall homepage data. The distribution derived from the unbiased recommendations is notably flatter, indicating a reduced popularity bias. In contrast, the overall data shows a heavy concentration of popularity among a small set of highly ranked content, reflecting a long-tail distribution often seen in popularity-driven systems.

To further quantify the difference in popularity concentration, we computed the Gini coefficient for both distributions. The Gini coefficient is a standard measure of inequality, where 0 indicates perfect equality and 1 indicates maximum inequality. The unbiased recommendation list yielded a Gini coefficient of 0.203, while the overall (biased) popularity list exhibited a significantly higher value of 0.494. This stark contrast confirms that the unbiased system distributes attention more evenly across content, whereas the biased system concentrates popularity in a small subset of items. This aligns with our earlier observation from the normalized distribution curves and provides quantitative support for the reduced popularity bias of the unbiased approach.

\subsection{Leveraging Unbiased Exploration Data for Candidate Generation}

Considering engineering implementation efforts and potential gain of using unbiased, a candidate retrieval enhancement online experiment on a large scale recommendation system has been conducted, which results in significant improvement on key user engagement metrics.

To address the bias introduced by traditional engagement-driven recallers, we developed an \textbf{Unbiased Co-Occurrence Recaller} that utilizes interaction signals collected from a uniform exploration container deployed in production. This container presents randomized, position-independent content, enabling the extraction of more reliable and unbiased preference signals from user interactions. To enlarge the impact of unbiased interaction, the behavior on whole page is utilized to create Co-Occurrence data. 

\subsubsection{Recaller Construction}

The Unbiased Co-Occurrence Recaller is constructed using an offline batch processing pipeline that computes pairwise co-engagement statistics from exploration-based interactions. Specifically, for each pair of titles $(A, B)$, we compute a normalized co-view metric based on user engagement:

\begin{itemize}
  \item Let $u$ be a user who \textit{viewed title B from the exploration container}, and who also has \textit{title A} in their historical viewing data.
  \item We aggregate the \textbf{user engagement on title B} across all such users $u$ who have also watched A.
  \item This defines a directional association: ``Given A in the watch history, B is likely to be relevant,'' based on unbiased exposure.
\end{itemize}

Only co-occurrence pairs exceeding a minimum interaction threshold are retained to ensure statistical significance. The resulting co-occurrence table captures a directed similarity signal between items that is minimally influenced by popularity or positional bias, as the source interactions are derived from randomized exposure.

\subsubsection{Candidate Retrieval at Serving Time}

At serving time, the Unbiased Co-Occurrence Recaller is used for candidate generation as follows:

\begin{enumerate}
  \item \textbf{History Extraction}: Retrieve the set of recently watched titles $\{A_1, A_2, ..., A_n\}$ from the user’s viewing history.
  \item \textbf{Lookup}: For each $A_i$, query the top-$K$ associated titles $\{B_1, ..., B_k\}$ using the precomputed co-occurrence table.
  \item \textbf{Aggregation}: Merge all associated $B_i$ results across all $A_i$.
\end{enumerate}

This candidate set is used in  \textit{home screen recommendations} where unbiased and diverse suggestions are essential for maintaining long-term user engagement.

\subsubsection{Results and Discussion}

The experiment was conducted on a large-scale streaming platform \ifanonymous \else with millions of active users\fi, evaluating the downstream impact of integrating the \textit{Unbiased Co-Occurrence Recaller} into the homepage recommendation pipeline. The control variant relied solely on the existing engagement-driven recallers, while the treatment variant augmented candidate generation with titles surfaced from unbiased exploration data collected via the \ifanonymous dedicated \else “Something Completely Different” \fi row.

As shown in Table~\ref{utilization}, the inclusion of the unbiased recaller led to a substantial lift in user engagement: a +0.94\% increase in key user engagement metric, statistically significant with a $p-value < 0.001$. Although the lift appears modest, such an improvement represents a substantial gain at the scale of millions of active users and translates into meaningful downstream benefits, including increased opportunities for advertising revenue. This result validates that signals extracted from uniformly randomized content exposure are not only cleaner in terms of bias but also highly actionable when reinjected into the recommendation stack.

\begin{table}[ht]
\caption{\textbf{Impact of Unbiased Co-Occurrence Recaller on Homepage Recommendation}}
\centering
\begin{tabular}{lcc}
\hline
\textbf{Treatment} & \textbf{User Engagement Lift} & \textbf{p-value} \\
\hline
Control (no extra recaller) & – & – \\
Unbiased Recaller (treatment) & \textbf{+0.94\%} & $ p < 0.001 $ \\
\hline
\end{tabular}
\label{utilization}
\end{table}

The improvement confirms that unbiased interaction signals, derived from behaviorally informed exposure, can enhance recommendation quality. This also demonstrates the broader utility of unbiased data beyond fairness or offline evaluation: it can directly contribute to measurable business outcomes when applied to candidate generation.

Notably, this gain was achieved without requiring complex changes to ranking models or personalization logic, highlighting the modular and scalable nature of the proposed exploration and data collection strategy. The results underscore that with the right UI placement and behavioral cues, unbiased data can be collected at scale and converted into meaningful user impact.

\subsection{Broader Applications of Unbiased Data}

Unbiased interaction data has far-reaching utility across the entire recommendation pipeline. In offline evaluation, such data provides a more reliable basis for benchmarking model performance, reducing overestimation caused by exposure bias in logged feedback \cite{li2011unbiased}. It can also support more accurate counterfactual estimation, enabling safer model iteration and validation \cite{liu2020general}.

Beyond evaluation, embedding learning benefits from unbiased exposure, as it avoids overfitting to popular items and enables more balanced representation learning across the item catalog \cite{liu2023bounding}.

Unbiased data can also be directly integrated into the ranking stage. For example, it can be used to pre-train models on more representative user-item distributions, or to fine-tune ranking objectives that correct for presentation bias\cite{bonner2018causal}. In large-scale deployments, unbiased datasets can even serve as a foundation for training new recommendation models from scratch-facilitating experimentation, fairness interventions, and algorithmic transparency.

Recognizing the importance of unbiased data, some platforms have gone as far as open-sourcing randomized datasets to encourage research progress in this area \cite{gao2022kuairand}. Our approach contributes to this growing ecosystem by making the collection of unbiased data operationally feasible and scalable in production systems.

\section{Conclusion and future work}

In this paper, we introduced a practical approach for exploration in recommender systems that effectively balances the tradeoff between high-quality data collection and short-term user engagement. By conditioning the delivery of exploratory content on empirically defined low-cost, high-reach scroll-depth regions, our approach enables randomized exposure at scale without compromising key performance metrics. The \ifanonymous dedicated \else “Something Completely Different” \fi row, deployed on a large-scale streaming platform, demonstrates that behaviorally informed placement can facilitate the collection of high-quality, unbiased interaction data in production environments.

Looking ahead, several promising directions emerge. First, rather than relying on a fixed scroll-depth row to infer exploration readiness, we plan to investigate more dynamic, model-driven signals. For example, user behaviors such as repeatedly scrolling right within a row, or completing several consecutive view sessions, may indicate a higher likelihood of receptiveness to novel content. Capturing this real-time intent could enable even more targeted and context-aware exploration.

Second, we aim to expand the scope of exploration from the program level to finer-grained content attributes such as genres, actors, or themes. Understanding user preferences at these dimensions could unlock more meaningful personalization and support use cases like cold-start item recommendation and taste discovery.

Finally, we plan to further leverage the collected unbiased data to train fairer and more robust recommendation models. This includes improving recall quality, enhancing personalization for underserved audiences, and supporting offline evaluation frameworks that better reflect real-world distributional shifts.

\bibliographystyle{IEEEtran}
\bibliography{myref}
\end{document}